**TITLE:**

Optimizing IMPULSED Acquisition Protocols for Clinical 3T Scanners Through Bayesian Experimental Design

**AUTHORS:**

Yan Dai B.S.[1], Xun Jia Ph.D.[2], Todd Aguilera M.D., Ph.D.[1], Kai Jiang Ph.D.[1], Arely Perez Rodriguez B.S.[1], Isabelle Vanhaezebrouck D.V.M.[1], Jie Deng Ph.D.[1]

[1] Department of Radiation Oncology, University of Texas Southwestern Medical Center, TX, USA

[2] Department of Radiation Oncology and Molecular Radiation Sciences, Johns Hopkins University, MD, USA

**CORRESPONDENCE:**

Jie Deng, Ph.D.

214-645-5140

Department of Radiation Oncology,

University of Texas Southwestern Medical Centre, 2280 Inwood Rd, Dallas.

Email: Jie.Deng@UTSouthwestern.edu

## Abstract

**Purpose:**

To optimize diffusion MRI acquisition protocols for IMPULSED (Imaging Microstructural Parameters Using Limited Spectrally Edited Diffusion) model at clinical 3T scanner using Bayesian experimental design, enabling accurate cellular-scale parameter estimation under realistic scan time and scanner hardware constraints.

**Methods:**

Expected Information Gain (EIG) was used as the optimization objective to maximize the information content of acquired measurements for IMPULSED model fitting. Bayesian optimization with Gaussian process surrogates efficiently searched the high-dimensional acquisition parameter space, including pulse types (PGSE, OGSEn1, and OGSEn2), diffusion times, and b-values. Optimized protocols were systematically evaluated against a heuristically designed baseline protocol through simulation studies assessing classification accuracy and parameter estimation performance across SNR levels of 5-40. Robustness to optimization assumptions was examined by varying prior distributions and assumed SNR. In-vivo validation was performed using canine tumor data acquired at 3T.

**Results:**

The optimized protocol eliminated OGSEn2 acquisitions, concentrated measurements at high b-values (maximum 2000 s/mm² for PGSE, 1220 s/mm² for OGSEn1), employing concurrently optimized diffusion timing. Compared to the baseline protocol, the optimized design achieved superior classification accuracy for distinguishing cell populations and reduced parameter estimation error across biologically relevant parameter ranges at various SNRs. Performance advantages were consistent across diverse optimization scenarios, demonstrating robustness to prior knowledge and noise assumptions. In-vivo parameter maps showed substantially improved quality and smoothness.

**Conclusion:**

Bayesian optimization substantially improves IMPULSED acquisition design for clinical 3T scanners. This principled, algorithm-agnostic framework enables accurate diffusion MRI cytometry under clinical constraints, with potential applications to tumor characterization and treatment monitoring.

## 1. Introduction

Diffusion magnetic resonance imaging (dMRI) characterizes tissue microstructure by measuring signal attenuation caused by water molecule displacement under diffusion-weighted gradients(1,2). Quantitative parameters are obtained by fitting biophysical models to the acquired signals, providing imaging biomarkers about cell microenvironment(3-6). The accuracy and robustness of these biomarkers depend on both measurement quality, determined by acquisition strategy and signal-to-noise ratio (SNR), and the geometry of the inverse problem underlying parameter fitting(7,8). While considerable effort has focused on improving parameter estimation algorithms(9-14), the design of the acquisition protocol itself remains a critical yet often under-optimized factor. Most clinical dMRI protocols are still designed heuristically rather than through principled experimental design(15-18), which can limit parameter identifiability for complex diffusion models.

Experimental design optimization generally involves defining a target function and searching for acquisition parameters that maximize this objective. Within diffusion MRI, experimental design optimization has been investigated for several commonly used signal models such as the apparent diffusion coefficient (ADC) and intravoxel incoherent motion (IVIM) models. Existing methods typically optimize the selection of b-values under scan-time constraints using target functions include variance-based metrics such as parameter estimation error(19-25), fisher information matrix(26), or the Cramér–Rao lower bound(27-29), and task-based objectives such as maximizing class distinguishability(30,31) or preserving linearity between estimated parameters and ground truth values(32-34). Because analytical gradients with respect to acquisition parameters are often unavailable or computationally intractable, these objectives are typically treated as black-box functions and optimized using numerical strategies such as brute-force search(21-23,25,29,30,32), genetic algorithms(19,20,24), or surrogate modeling approaches(35,36).

Critically, these strategies rely on specific parameter estimation algorithms, limiting generalizability, and approximate uncertainty using simplified lower-order metrics, which may be inadequate for complex qMRI models. Consequently, they are not directly extensible to modern diffusion models with high-dimensional parameter spaces and strongly ill-conditioned inverse mappings, such as the Imaging Microstructural Parameters Using Limited Spectrally Edited Diffusion (IMPULSED) model(37) considered in this work.

IMPULSED is an advanced diffusion MRI framework that models water displacement in intracellular and extracellular compartments across multiple effective diffusion times using a combination of pulsed gradient spin echo (PGSE) and oscillating gradient spin echo (OGSE) waveforms(38,39). This approach enables estimation of cellular-scale parameters such as cell size and intracellular volume fraction, which have shown specificity for characterizing radiation-induced biological changes in both preclinical and clinical studies. Although successful implementations have been demonstrated on higher-field scanners (e.g., 4.7T, 7T) using empirically selected diffusion times and evenly distributed b-values(17,37,40), translating the IMPULSED model to widely used clinical 3T systems remains challenging. Reduced gradient strength limits the achievable diffusion time range, and clinically feasible scan times restrict the ability to compensate for low SNR through signal averaging. In addition, the highly nonlinear nature of the IMPULSED model leads to an ill-conditioned parameter estimation problem. These challenges necessitate optimized acquisition designs that maximize the information content of the measurements. However, experimental design for IMPULSED model fitting involves a high-dimensional parameter space and an extremely ill-conditioned parameter-fitting landscape, making many existing dMRI experimental design optimization approaches inapplicable.

In this work, we formulate IMPULSED protocol design optimization as an information-theory-based problem, using Expected Information Gain (EIG) as the target function(41,42). To our knowledge, this is the first study to apply information-theoretic experimental design to diffusion MRI acquisition. Bayesian optimization with Gaussian process surrogate modeling is employed to efficiently search the high-dimensional acquisition parameter space. This framework directly maximizes the information content acquired and independent of downstream parameter estimation algorithms, enabling general application across different fitting strategies. We systematically analyzed the effectiveness and robustness of the proposed experimental design optimization framework for the IMPULSED model under realistic clinical constraints.

**2. Methods**

2.1 IMPULSED Model

The IMPULSED model is a two-compartment framework assuming isotropic water diffusion both inside and outside sphere-shaped cells with impermeable membranes. Conceptually, it connects the observed diffusion signal $S$ with contributions from restricted intracellular and extracellular diffusion. The signal is expressed as a weighted sum of intracellular and extracellular contributions: $S = V_{in} \cdot S_{in} + (1 - V_{in}) \cdot S_{ex}$, where $V_{in}$ is the intracellular volume fraction, and $S_{in}$ and $S_{ex}$ are signals from the intra- and extracellular compartments, respectively. The IMPULSED model describes diffusion signal decay arising from PGSE, OGSEn1, and OGSEn2 acquisition pulses, where $S_{ex}$ is expressed as $S_{ex} = \exp(-b(D_{ex0} + \beta_{ex} \cdot f))$, with $f$ = 0, 1, 2 for PGSE, OGSEn1 and OGSEn2, respectively, and $S_{in}$ is expressed as Eq (1) for PGSE and Eq (2) for OGSEs.

$$S_{in} = \exp\left(-2\left(\frac{\gamma G}{D_{in}}\right)^2 \sum_n \frac{B_n}{\lambda_n^2} \{\lambda_n D_{in} \delta - 1 + \exp(-\lambda_n D_{in} \delta) + \exp(-\lambda_n D_{in} \Delta)(1 - \cosh(\lambda_n D_{in} \delta))\}\right),$$ Eq. 1

$$S_{in} = \exp\left(-2(\gamma G)^2 \sum_n \frac{B_n \lambda_n^2 D_{in}^2}{(\lambda_n^2 D_{in}^2 + 4\pi^2 f^2)^2} \left\{\frac{(\lambda_n^2 D_{in}^2 + 4\pi^2 f^2)}{\lambda_n D_{in}} \left[\frac{\delta}{2} + \frac{\sin(4\pi f \delta)^2}{8\pi f}\right] - 1 + \exp(-\lambda_n D_{in} \delta) \right. \right.$$
$$\left. \left. + \exp(-\lambda_n D_{in} \Delta)(1 - \cosh(\lambda_n D_{in} \delta))\right\}\right).$$ Eq. 2

Microstructural parameters, including cell radius ($R$), intracellular volume fraction ($V_{in}$), intracellular diffusion coefficient ($D_{in}$), and extracellular diffusion parameters ($D_{ex}$, $\beta_{ex}$) can be estimated by fitting the IMPULSED model. Here, $\gamma$ denotes the gyromagnetic ratio; $\delta$ and $\Delta$ represent the duration of diffusion gradient pulse and the time interval between the two diffusion gradient pulses around the 180° refocusing pulse, respectively; and $G$ denotes the diffusion gradient amplitude. $\lambda_n$ and $B_n$ are structure-dependent parameters determined by the cell radius $R$. This model enables measurement of diffusion behavior across a wide range of $t_{\text{diff}}$ through the combined use of PGSE, OGSEn1, and OGSEn2 waveforms. For PGSE, the effective diffusion time is given by $t_{\text{diff}} = \Delta - \delta/3$, whereas for OGSE, $t_{\text{diff}} = \delta/4f$, where $f$ denotes the oscillating frequency. The $b$-value commonly used in dMRI does not explicitly appear in this representation, but is instead decomposed into $g, \delta$ and $\Delta$, such that $b = \gamma^2 \int_0^{+\infty} dt \left| \int_0^t dt' g(t') \right|^2$, where $g(t)$ denotes the diffusion gradient waveform with amplitude $G$. Thus, given fixed $\delta$ and $\Delta$, the $b$-value can be considered as a one-to-one mapping of $G$.

2.2 Tunable Protocol Parameters for Experimental Design Optimization

The tunable parameters for each acquisition include the pulse type (PGSE, OGSEn1, or OGSEn2) and the corresponding timing parameters ($\delta$ and $\Delta$), which determine the effective diffusion time $t_{\text{diff}}$. In addition, the $b$-values, or equivalently, the diffusion gradient amplitude $G$ for a given pulse type, need to be determined. To minimize relaxation-related effects, the echo time (TE) and repetition time (TR) were kept the same for all pulse types (n=0, 1, 2).
In our implementation on a scanner (3T Philips Achieva MR system), we adopted timing parameters following the baseline protocol applied by Xu et al. , with a gradient pulse duration of 41 ms, a pulse gap of 10 ms between gradient pulses, yielding the same TE = 103 ms. The TR was matched to their reported value of 4500 ms. The maximum gradient amplitude $G$ was set to 8×10⁻⁴ gauss/μm, according to the hardware constraints of the 3T system.

The bounds for timing parameters were determined to ensure physically realizable protocols. First, the lower bound of $\delta$ for each pulse type was constrained to ensure the maximum achievable b-value exceeded 100 s/mm²: $\delta_{\text{PGSE}} \geq 10$ ms, $\delta_{OGSEn1} \geq 19$ ms, and $\delta_{OGSEn2} \geq 30$ ms. Given these minimum $\delta$ values, the upper bound of $\Delta$ for each sequence type was derived as: $\Delta_{max} = 2 \times$ pulse duration $-$ pulse gap $- \delta_{min}$, yielding $\Delta_{PGSE} \leq 82$ ms, $\Delta_{OGSEn1} \leq 73$ ms, and $\Delta_{OGSEn2} \leq 62$ ms. The lower bound of $\Delta$ was set to ($\delta_{\min}$ + pulse gap) to account for the duration of the 180° refocusing pulse. For each specific $\Delta$ value during optimization, the upper bound of $\delta$ was dynamically determined as $min(\Delta -$ Pulse Gap, $2 \times$ pulse duration $-$ pulse gap $- \Delta)$.

To avoid non-Gaussian diffusion effects associated with excessively high $b$-values, and considering the highest $b$-value 1800 s/mm² applied by by Xu et al, the $b$-value was capped at 2000 s/mm². Following the acquisition design of Xu et al., a complete IMPULSED acquisition protocol was limited to 13 nonzero b-value measurements plus 3 zero-b measurements for each pulse type, corresponding to a total scan time of approximately 4 minutes on our 3T scanner. This consistent protocol structure enables direct comparison between our optimized acquisition schemes and the baseline protocol.

2.3 Target Function: Expected Information Gain

EIG, derived from Bayesian experimental design theory, was used as the target function to quantify the amount of information obtained from a given experimental design $d$. EIG is defined in Eq (3).

$$EIG(d) = E_{P(S_m|d)} \left[ E_{P(\theta|S_m, d)}[\ln P(\theta|S_m, d)] - E_{P(\theta|d)}[\ln P(\theta|d)] \right], \quad \text{Eq. 3}$$

where $S_m = S_{m,1}, S_{m,2}, \ldots, S_{m,K}$ represents the measured dMRI signals for a voxel with microstructural parameters $\theta = (R, V_{in}, D_{in}, D_{ex}, \beta_{ex})$ acquired using experimental design $d$. EIG measures the expected reduction in entropy from the prior to the posterior distributions of the model parameters.

EIG was evaluated numerically by approximating the required expectations over both the parameter space and the measurement noise space. The prior distribution $P(\theta|d) = P(\theta)$ was assumed to be uniform within physiologically-relevant ranges(37): $R \in [2, 12]$ µm, $D_{in} \in [0.2, 2]$ µm²/ms, $V_{in} \in [0.2, 0.8]$, $D_{ex} \in [0.2, 2]$ µm²/ms and $\beta_{ex} \in [0, 6]$ µm². To approximate the expectation over the prior, a total of $6 \times 4 \times 6 \times 4 \times 4$ parameter samples were generated using grid-based sampling within this space, with denser sampling applied for cell radius $R$ and intracellular volume fraction $V_{in}$. The posterior distribution was computed as $P(\theta|S_m, d) = P(\theta|S_m) = P(S_m|\theta)P(\theta)/P(S_m)$. Assuming additive Gaussian noise in the measured signals, which provides a good approximation to Rician noise at $SNR > 3$, then $P(S_m|\theta, d) = \prod_{i=1}^{K} \mathcal{N}(S_{m,i}; f_i(\theta, d), \sigma_i)$, where $\mathcal{N}(.)$ represents a Gaussian distribution with mean $f_i(\theta, d)$ and noise level $\sigma_i$. Here, $f_i(\theta, d)$ represents the noiseless IMPULSED signal for the $ith$ acquisition setting under design $d$. The expectation over measurement noise was estimated using Monte Carlo sampling. For each parameter sample, 100 noisy signal realizations $P(S_m|\theta, d)$ were generated by repeatedly adding Gaussian noise to the noiseless signal at an $SNR = 20$, representative of typical clinical imaging conditions. The EIG was then computed by averaging over all simulated measurements and parameter samples. By maximizing this target function, the optimal acquisition design $d^*$ was identified.

2.4 Bayesian Optimization Procedure

The acquisition protocol design was formulated as a 32-dimensional continuous optimization problem as Bayesian Optimization input. Specifically, a protocol consisting of 13 individual measurements was encoded as a vector with $6 + 13 \times 2$ elements. The 6 shared parameters correspond to the timing variables $\delta$ and $\Delta$ for each of the three diffusion gradient waveform types (PGSE, OGSEn1, or OGSEn2). Each of the 13 acquisitions contributes 2 parameters: the diffusion gradient pulse type indicator that specifies whether the acquisition uses PGSE, OGSEn1, or OGSEn2. and its corresponding b-value. To enable optimization within a continuous parameter space, the gradient pulse type indicator was represented by a continuous indicator variable in the interval [0,1]. During protocol evaluation, this variable was mapped to the three discrete waveform types through interval partitioning: values in [0, 1/3) correspond to PGSE, values in [1/3, 2/3) correspond to OGSEn1, and values in [2/3, 1] correspond to OGSEn2. Each dimension of the vector was normalized from its physical bounds to the unit interval [0,1], enabling efficient exploration of the parameter space by the Gaussian process surrogate model.

Gaussian process (GP)–based Bayesian optimization was employed to efficiently explore the EIG as a function of acquisition parameters. A Matérn kernel was selected to model the structure of the EIG surface, enabling a flexible surrogate representation of the underlying black-box objective. The optimization process was initiated with 300 protocol evaluations generated via quasi-random Sobol sampling, followed by 500 sequential optimization iterations. At each iteration, new candidate acquisition designs were selected by maximizing the log expected improvement acquisition function, balancing exploration and exploitation. This GP-based Bayesian optimization framework follows standard practice in black-box experimental.

2.5 Evaluation

The experimental designs were evaluated comparing between an optimized acquisition protocol obtained by maximizing EIG under a uniform prior distribution with an assumed SNR of 20 (denoted *Optimized SNR 20*) and the protocol applied in previous literature(17). as a baseline (*Baseline*). The evaluation comprised two main components. For simulation study, we first assessed the performance advantage of the optimized acquisition protocols relative to conventional baseline protocols in terms of both classification and parameter estimation accuracy under different noise conditions. Then, we investigated the behavior of the optimization process itself by analyzing how the structures of the optimized protocols varied under different optimization assumptions, including assumed noise level and prior parameter distribution. Additionally, we performed scans with both the baseline and optimized protocols on a canine patient, and the resulting parameter maps were compared qualitatively.

2.5.1 Simulation Study

2.5.1.1 Classification

**Data generation**
Synthetic datasets were generated to evaluate the classification performance of different IMPULSED acquisition protocols. Two biologically relevant microstructural variations were simulated: immune-cell infiltration, affecting cell radius $R$, and tumor apoptosis, affecting intracellular volume fraction $V_{in}$. For each experiment, only one parameter ($R$ or $V_{in}$) was varied across the same ground truth values used for prior sampling, yielding six distinct 'tissue states'. All other parameters were fixed at physiologically representative

values(37) ($R = 7 \mu m$, $v_{in} = 0.6, D_{ex} = 1.1 \, \mu m^2/ms$, $D_{in} = 0.9 \, \mu m^2/ms$ and $\beta_{ex} = 2 \, \mu m^2$). For each parameter setting, 500 independent noisy signal realizations were generated.

Noisy signals were simulated using a Rician noise model. Noise-free signals were first generated from the IMPULSED model for each acquisition protocol, after which complex Gaussian noise with standard deviation σ was added to the real and imaginary components, followed by magnitude reconstruction. Four noise levels corresponding to SNR of 5, 10, 20, and 40 were evaluated, where $SNR = 1/\sigma$.

**Classification Algorithm**

Classification was performed using a Random Forest classifier. Each noisy signal vector, consisting of signal intensities from all acquisitions using a given protocol, served as the input feature. Integeral Class labels (0-5) were assigned, corresponding to increasing ground truth values of $R$ or $V_{in}$.

**Evaluation Metric**

Classification performance was evaluated using 10-fold cross-validation. For each protocol and SNR level, the mean classification accuracy and standard deviation across folds were computed. Principal component analysis (PCA) was additionally applied to visualize class separability in the signal space.

2.5.1.2 Quantification

**Data Generation**

A synthetic dataset containing 100,000 parameter combinations was generated by uniformly sampling microstructural parameters within predefined physiologically relevant ranges. For each parameter set, noisy dMRI signals were simulated using a Rician noise model at SNRs of 5, 10, 20, and 40, following the procedure described above. All signals were normalized using b0 acquisitions prior to parameter fitting.

**Quantification Algorithm**

Parameter estimation was performed using nonlinear least-squares (NLLS) optimization implemented with the Trust Region Reflective (TRF) algorithm. To improve fitting stability, $D_{in}$ was fixed at 1.58 $\mu m^2/ms$ and $\beta_{ex}$ at 2, leaving three free parameters ($R$, $V_{in}$ and $D_{ex}$)(17). Parameter bounds were set to $R \in [0.2, 15] \, \mu m, V_{in} \in [0, 1.0]$, and $D_{ex} \in [0, 3] \, \mu m^2/ms$. Initial parameter values were randomly sampled within the predefined physiologically-relevant ranges for each optimization run.

**Evaluation Metric**

Quantification accuracy was evaluated using root mean square error (RMSE), normalized by the corresponding upper bounds of the fitting constraints ($R$: 15 μm, $V_{in}$: 1.0, $D_{ex}$: 3 $\mu m^2/ms$). To examine spatial variation in estimation accuracy, the ($R$,$V_{in}$) parameter space was discretized into a 5×5 grid using bin edges

$R = [2.0, 4.0, 6.0, 8.0, 10.0, 12.0]$ μm and $V_{in} = [0.2, 0.32, 0.44, 0.56, 0.68, 0.8]$, resulting in 25 distinct biophysical regimes spanning from small cells with low intracellular volume to large cells with high intracellular volume. Parameter-specific RMSE values were computed independently within each bin.

2.5.1.3 Ablation Analysis

Two additional experiments were conducted to isolate the contributions of specific experimental design elements. First, an optimized protocol derived by maximizing EIG under the same uniform prior distribution but assuming a higher SNR of 40 (*Optimized SNR 40*) was evaluated to assess the sensitivity of the optimization outcome to noise assumptions. Second, a modified baseline protocol (*Baseline2*) was constructed by adopting the optimized $t_{\text{diff}}$ from the *Optimized SNR 20* protocol while retaining the *Baseline* b-value distribution.

2.5.1.4 Dependence of Optimized Protocols on Prior Distributions and SNR

To examine how optimization outcomes depend on prior assumptions and noise conditions, acquisition protocols were optimized under 13 experimental configurations comprising nine prior distributions and four SNR levels.

The prior distributions represented varying degrees of biological knowledge about underlying cell populations. Three configurations for cell radius $R$ (uniform (UR), small-cell-biased (SR), and large-cell-biased (LR)) were combined with three configurations for intracellular volume fraction $V_{in}$ (uniform (UVin), small-$V_{in}$-biased (SVin), and large-$V_{in}$-biased (LVin)), this yields 8 non-uniform prior combinations UR-SVin, UR-LVin, SR-UVin, SR-SVin, SR-LVin, LR-UVin, LR-SVin, and LR-LVin, aside from UR-UVin. Each prior distribution was modeled using a linear probability density $P(x) \propto (1 - ratio) \cdot u + ratio$, where $u = (x - x_{min})/(x_{max} - x_{min})$. Ratio of 10 and 0.1 were used to generate the small-biased and large-biased priors, respectively. Details of prior distributions are shown in **Figure S1**. For prior comparisons, the SNR was fixed at 20. Further, to assess SNR assumption dependence, additional optimizations were performed at SNR levels of 5, 10, and 40 using the uniform prior (UR-UVin).

2.5.2 In-vivo Study

To evaluate the optimized experimental design in vivo, diffusion MRI data were acquired from a canine tumor model on a clinical 3T MRI scanner (Philips Achieva). The subject had a squamous cell carcinoma located in the temporal region dorsal to the left eye. Both the *Baseline* and *Optimized SNR 20* protocols were applied. A tumor region of interest (ROI) was manually segmented, and the same NLLS algorithm was used to estimate the parameters $R$, $V_{in}$, and $D_{ex}$ for each protocol. Parameter maps were generated for qualitative comparison of estimation stability.

**3. Results**

3.1 Experimental Optimization Results

The detailed acquisition parameters of the optimized experimental design under the uniform prior distributions at SNR of 20 (*Optimized SNR 20*), together with the baseline protocol (Baseline), are shown in **Table 1**. The *Optimized SNR 20* protocol exhibits several distinctive features compared to the baseline protocol. Most notably, the optimization eliminated the OGSEn2 pulse type entirely, reallocating those measurements to PGSE and OGSEn1 sequences. For OGSEn1, the optimized protocol concentrated all five measurements at the maximum achievable b-value of 1220 s/mm², utilizing the maximum gradient duration ($\delta$ = 41 ms, $\Delta$ = 51 ms) identical to the baseline protocol, thereby maximizing the effective diffusion time for this pulse type.

The PGSE component showed a similar sampling strategy. The optimized protocol allocated eight measurements to PGSE, with sparse sampling at intermediate b-values (500-1000 s/mm²) and dense sampling at the maximum b-value of 2000 s/mm². However, unlike OGSEn1, the PGSE timing parameters ($\delta$ = 33 ms, $\Delta$ = 59 ms) resulted in a relatively shorter effective diffusion time compared to that for *Baseline*. The convergence behavior of the Bayesian optimization process, along with the EIG values for optimizations is shown in **Figure S2**.

3.2 Simulation study: Superior Performance of Optimized Protocols

3.2.1 Classification Performance

Classification accuracy for distinguishing cell populations based on different cell radius ($R$) and intracellular volume fraction ($V_{in}$) is shown in **Figure 1** across SNR levels of 5, 10, 20, and 40. For both parameters, the *Optimized SNR 20* protocol consistently achieved higher classification accuracy than the *Baseline* protocol across all SNR levels. Performance improvement was most pronounced at intermediate SNR levels and became less obvious at very high or very low SNRs, where all methods performed similarly well or poorly. PCA of signal vectors with varying $R$ and $V_{in}$ illustrated enhanced class separability. The *Optimized SNR 20* protocol produced substantially improved separation between classes compared with the *Baseline* protocol for both $R$ and $V_{in}$, yielding distinct and well-separated clusters in the dominant principal component space. The numerical results are shown in **Table 2**.

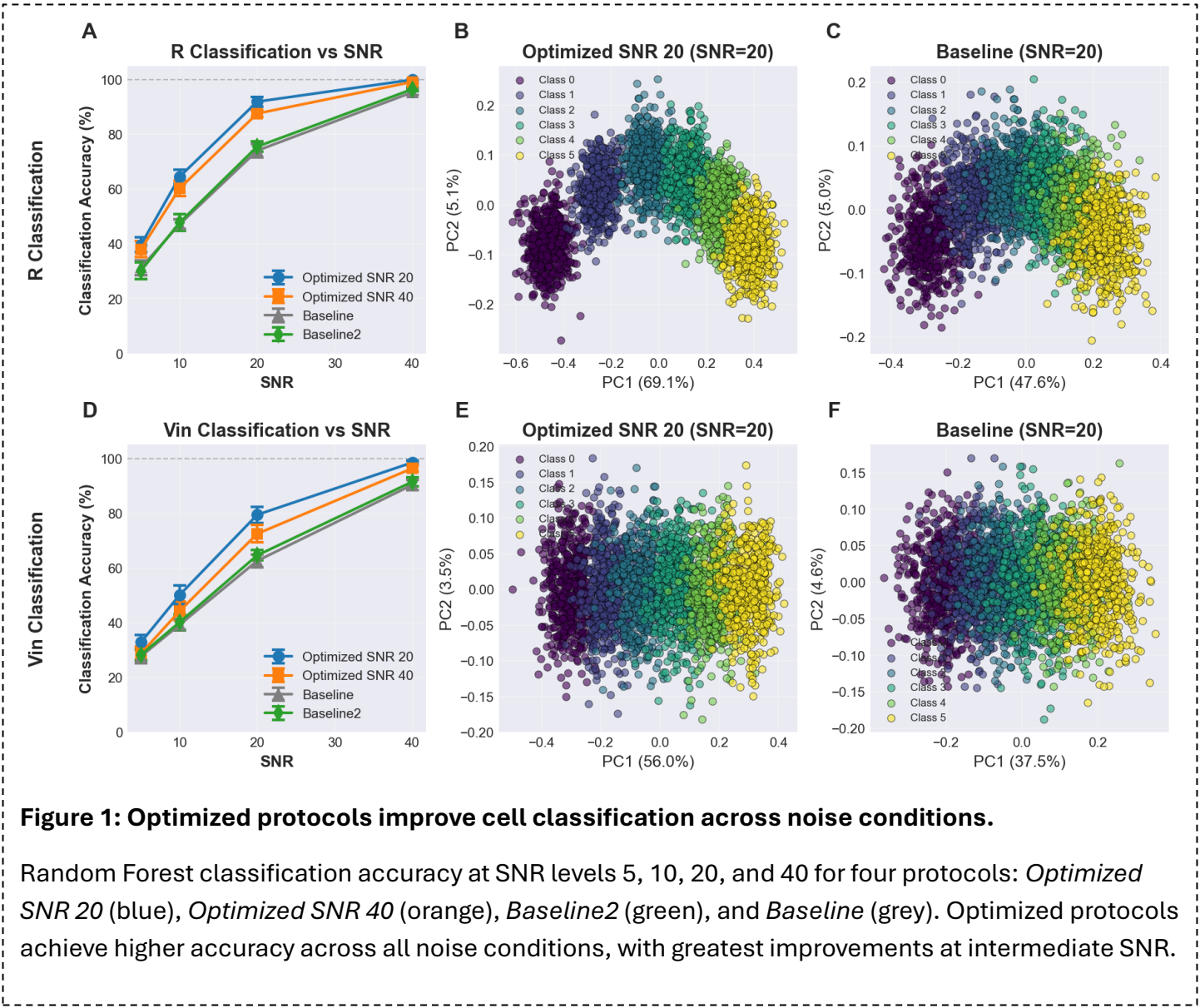

**Figure 1: Optimized protocols improve cell classification across noise conditions.**

Random Forest classification accuracy at SNR levels 5, 10, 20, and 40 for four protocols: *Optimized SNR 20* (blue), *Optimized SNR 40* (orange), *Baseline2* (green), and *Baseline* (grey). Optimized protocols achieve higher accuracy across all noise conditions, with greatest improvements at intermediate SNR.

3.2.2 Quantification Performance

**Figure 2** presents normalized RMSE for three biophysical parameters ($R, V_{in}, D_{ex}$) across SNR levels of 5, 10, 20, and 40. For all protocols and parameters, normalized RMSE decreased monotonically with increasing SNR. Across the entire noise spectrum, the optimized protocol *Optimized SNR 20* maintained a consistent performance advantage over the *Baseline* protocol for all parameters, demonstrating superior quantification accuracy and robustness to noise. The numerical results are shown in **Table 3**.

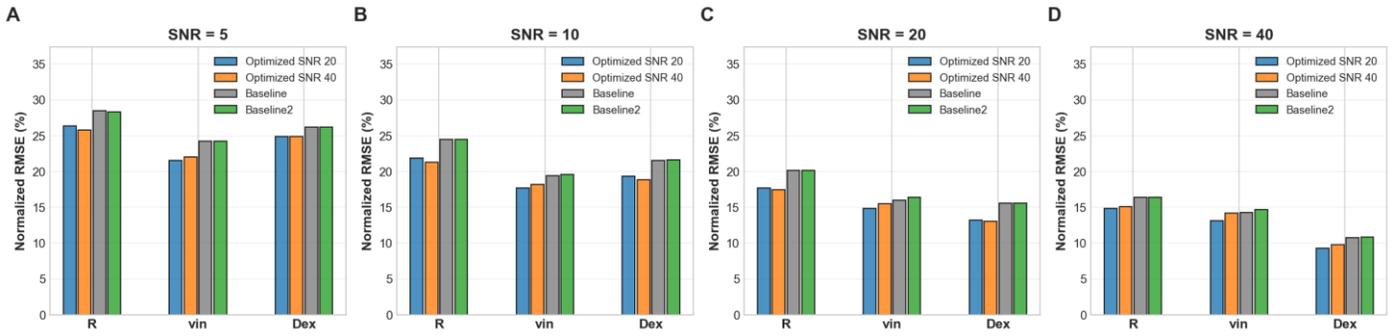

**Figure 2:** Optimized protocols reduce parameter estimation errors across noise conditions.

Normalized RMSE (% of parameter maximum) for 3-parameter fitting at SNR 5, 10, 20, and 40. Optimized protocols (blue, orange) substantially reduce RMSE compared to Baseline (grey) and Baseline2 (green).

Spatially resolved performance differences are shown in **Figure 3**, which presents 2D heatmaps of binned RMSE differences (*Optimized SNR 20* minus *Baseline*) for $R$ and $V_{in}$, quantification across SNR levels. Negative values (blue regions) indicate superior performance of the optimized protocol. Most of the regions exhibited superior baseline performance. For $R$ quantification, the greatest improvements were observed in the small-cell regime ($R$ < 8 μm) at higher intracellular volume fractions ($V_{in}$ > 0.5), with particularly strong benefits at SNRs of 10 and 20. For $V_{in}$ quantification, the optimized protocol demonstrated robust improvements across nearly the entire $(R, V_{in})$ space at SNRs of 10 and 20, with the strongest improvements occurring at larger $V_{in}$ regions.

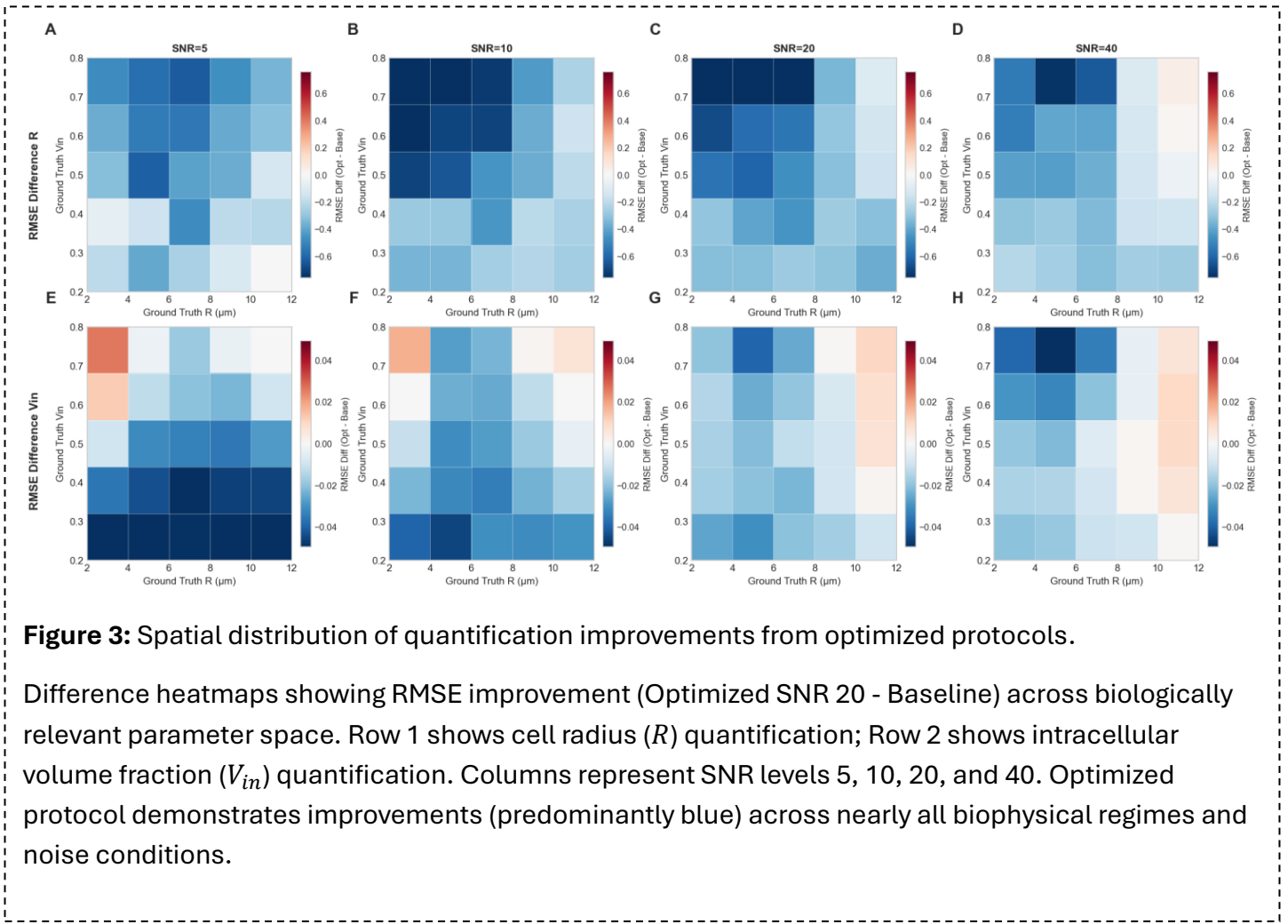

**Figure 3:** Spatial distribution of quantification improvements from optimized protocols.

Difference heatmaps showing RMSE improvement (Optimized SNR 20 - Baseline) across biologically relevant parameter space. Row 1 shows cell radius ($R$) quantification; Row 2 shows intracellular volume fraction ($V_{in}$) quantification. Columns represent SNR levels 5, 10, 20, and 40. Optimized protocol demonstrates improvements (predominantly blue) across nearly all biophysical regimes and noise conditions.

Collectively, **Figures 1-3** demonstrate that experimental design optimization provides higher quality data that substantially improve both classification and parameter estimation accuracy across diverse biophysical regimes and noise conditions.

3.3 Ablation Study

3.3.1 Impact of Diffusion Time and $b$-values

As shown in **Figure 1** the *Baseline2* protocol demonstrated slightly improved performance for classification of both $R$ and $V_{in}$, especially for $SNR \geq 10$. However, it remained inferior to the fully optimized protocol (*Optimized SNR 20*) across all SNR levels. For quantification (**Figure 2**), *Baseline2* achieved similar RMSE compared with the *Baseline* protocol across all SNRs for all parameters.

3.3.2 Impact of SNR level

As shown in **Figures 1** the *Optimized SNR 40* protocol demonstrated improved performance comparing with the *Baseline* protocol, and a worse performance comparing with the *Optimized SNR 20* protocol across all SNR levels. For quantification (**Figure 2**), the *Optimized SNR 40* protocol achieved similar RMSE as *Optimized SNR 40*. both optimized protocols substantially outperformed the Baseline and *Baseline* protocols across all SNR conditions, confirming the robustness and generalizability of Bayesian experimental design optimization for IMPULSED.

3.4 Impact of Assumptions on Optimized Acquisition Protocols

Optimized pulse sequence parameters across four SNR levels and nine prior distribution configurations are summarized in **Table 4** and **5**. The results reveal clear SNR-dependent and prior-dependent patterns for both b-value distributions and diffusion timing parameters across pulse types.

For PGSE acquisitions, b-value sampling was concentrated at high b-value region approaching the maximum achievable value (2000 s/mm²), with multiple acquisitions at the maximum b-value. As SNR increased, sampling expanded towards intermediate-to-high b-value, a pattern consistently observed across most prior distributions at SNR 20, indicating robustness to prior assumptions. The optimized PGSE diffusion time also showed an increasing trend with SNR, from $t_{\text{diff}} = 46.67$ ms at low SNR to $t_{\text{diff}} = 53.33$ ms at higher SNR level. Notably, $t_{\text{diff}}$ of PGSE exhibited sensitivity to prior distributions, with longer $t_{\text{diff}}$ favored under small-cell-biased priors. For OGSEn1 acquisitions, optimized protocols consistently concentrated measurements at the maximum achievable b-value (~1220 s/mm²) across all SNR levels and prior distributions. The $t_{\text{diff}}$ for OGSEn1 remained constant regardless of both SNR and prior distribution, consistently maintaining the maximum gradient duration ($\delta = 41\ ms$). OGSEn2 acquisitions were largely excluded from optimized protocols. At lower SNRs (5, 10, and 20), OGSEn2 was absent from all optimized protocols. It appeared only rarely across prior distributions at SNR 20 and emerged consistently only at the highest SNR (40).

3.5 In-vivo study: Improved Parameter Estimation Stability with Optimized Protocol

Visual comparison of parameter maps from *Baseline* and *Optimized SNR 20* were shown in **Figure 4**. Parameter maps derived from the optimized protocol (*Optimized SNR 20*) exhibited notably smoother spatial distributions across all three estimated parameters ($R, V_{in}, D_{ex}$) compared to the *Baseline* protocol. The optimized protocol substantially reduced the occurrence of extreme pixel values, which were present in the *Baseline* results.

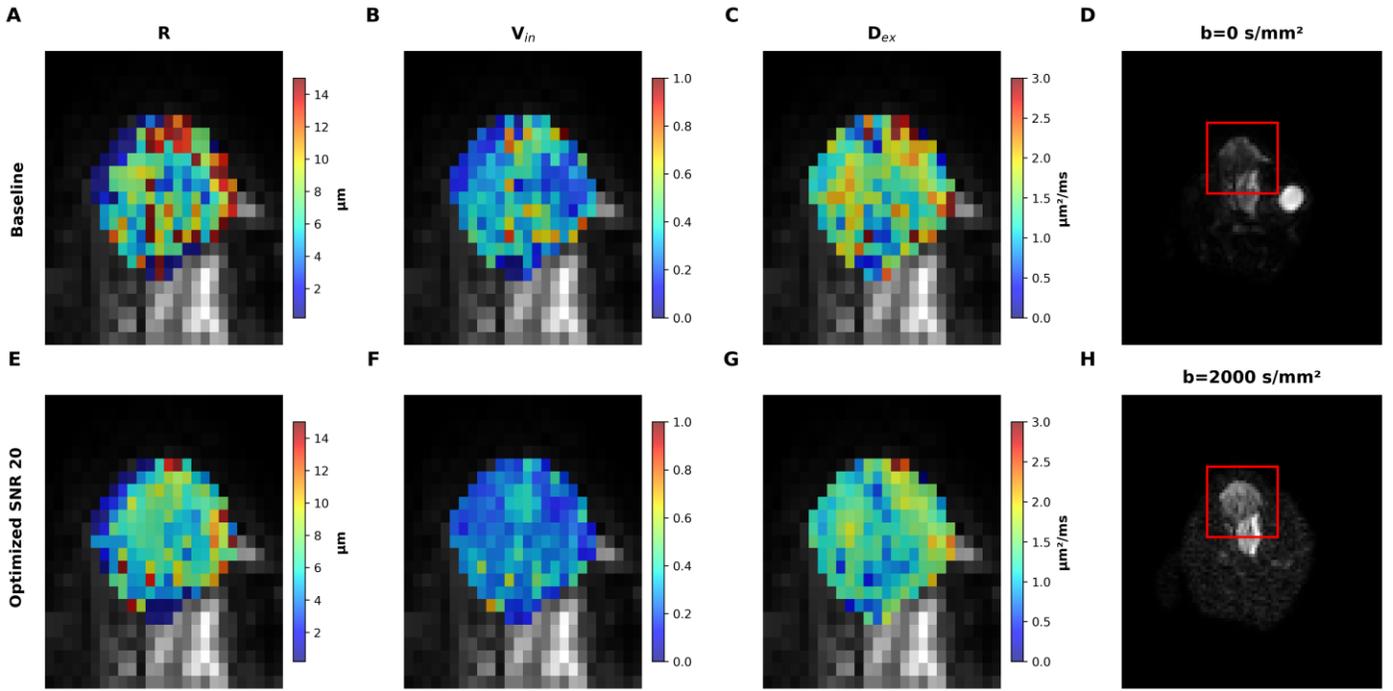

**Figure 4.** In vivo comparison of baseline and optimized protocols for IMPULSED parameter estimation in canine tissue.

Parameter maps derived from acquired in a canine model using two acquisition protocols. 1st row, 1~3 column: Results from the *Baseline* protocol. 2nd row, 1~3 column: Results from the SNR-*Optimized SNR 20* protocol. The estimated parameter maps are all overlaid on PGSE b=0 images within the tumor region of interest; Column 4 displays the full field-of-view PGSE diffusion-weighted images (top: b=0 s/mm², bottom: b=2000 s/mm²) with the red box indicating the zoomed region shown in columns 1-3; $R$: cell radius, $V_{in}$: intracellular volume fraction, $D_{ex}$: extracellular diffusivity; All parameter maps were estimated using nonlinear least-squares fitting with identical boundary constraints ($R$: $0.2 - 15$ μm, $V_{in}$: $0 - 1$, $D_{ex}$: $0 - 3.0$ μm²/ms). The optimized protocol yields visibly smoother parameter maps with reduced pixel-to-pixel variability and fewer extreme values, demonstrating improved estimation stability compared to the baseline protocol.

## 4. Discussion

4.1 Insights into Optimized Acquisition Strategies

Across optimization scenarios, the optimized protocols consistently concentrated measurements at high b-values. This pattern indicates that IMPULSED signals—particularly those acquired with OGSEn1 and OGSEn2—are highly sensitive to measurement noise, such that observed signal decay under clinical SNR conditions is effectively dominated by mono-exponential behavior. Consequently, only a subset of model parameters can be reliably identified in vivo. At SNR=40, OGSEn2 acquisitions with shorter $t_{diff}$ began to emerge, indicating partial recovery of additional parameter sensitivity as noise decreases. In contrast, OGSEn2 was systematically excluded at lower SNR levels. This reflects a trade-off between the sensitivity of high-frequency oscillating gradients to short-range cellular structure and their limited achievable b-values under hardware constraints. Under low SNR, the already-weak signal decay produced by low-b OGSEn2 acquisitions falls below the noise floor, contributing measurement variance rather than biophysical information. Concentrating acquisitions at high b-values therefore represents a pragmatic strategy to prioritize reliable estimation of identifiable parameters: optimized protocols prioritize reliable estimation of identifiable parameters rather than attempting to exhaustively characterize all aspects of the model.

The optimized diffusion timing parameters also exhibited systematic SNR-dependent patterns. As SNR increased, the optimized protocols expanded the range of effective diffusion times. This behavior is consistent with information-theoretic expectations: higher measurement precision allows acquisition strategies to exploit a broader range of contrast mechanisms.

An unexpected relationship was observed between the prior distribution and the optimized PGSE diffusion time. Priors biased toward smaller cell sizes resulted in longer optimized diffusion times, contrary to the conventional expectation that longer diffusion times primarily enhance sensitivity to larger structures. This effect may arise from interactions between the multi-compartment structure of the IMPULSED model and the information content at different diffusion timescales. One potential explanation is that for small cells, longer diffusion times may enhance the contrast between intracellular and extracellular compartments, thereby improving parameter identifiability despite probing spatial scales larger than the cell dimensions.

4.2 Relative Contributions of Acquisition Parameters

The ablation study using *Baseline2* protocol provided insights into the relative contributions of diffusion timing and b-value optimization. *Baseline2* adopted the optimized diffusion times from the *Optimized SNR 20* protocol while maintaining b-value distribution of *Baseline*. Although Baseline2 showed slight improvement over the Baseline protocol, it remained substantially inferior to the fully optimized protocol for classification tasks (**Figure 1**), indicating that b-value distribution plays a more dominant role than diffusion timing in determining measurement quality.

The influence of assumed SNR during optimization was evaluated by comparing Optimized SNR 40 and Optimized SNR 20 protocols. Despite structural differences between the two protocols, their performance across the tested SNR range was similar. This suggests that the EIG landscape contains multiple near-optimal solutions with distinct acquisition configurations that achieve comparable performance.

4.3 Practical Implications for Clinical Protocol Design

These results provide practical guidance for clinical protocol design. For typical clinical SNR levels (10–20), acquisition resources should primarily target PGSE and OGSEn1 measurements with intermediate-to-high b-values, while OGSEn2 contributes limited additional information and can be omitted to improve efficiency. OGSEn2 may only become beneficial at higher SNR (>40), where measurement precision allows the additional short-diffusion-time information to be effectively utilized. The robustness of optimized protocols' performance across SNR levels further suggests that a protocol optimized for SNR 20 can be reliably deployed across a range of clinical imaging conditions. Furthermore, the prior-dependent modulation of optimal PGSE diffusion times suggests that in settings where strong prior anatomical knowledge is available, protocol customization based on expected tissue characteristics could yield additional performance gains.

Note that in **Figure 1**, the Optimized protocols consistently outperformed the *Baseline* protocol, with the largest improvements observed at intermediate SNRs (10, 20). At low SNR (5), the performance advantage diminished because measurements were dominated by noise. Conversely, at high SNR (40), the *Baseline* protocol already achieved good performance, reducing the benefit of optimization. This SNR-dependent performance pattern indicates our optimized protocols will benefit the clinic application of IMPULSED model, where a typical SNR ranges at 10~20.

4.4 Quantification Challenges of the IMPULSED Model

The IMPULSED model involves a highly nonlinear mapping between microstructural parameters and diffusion signals, resulting in an ill-conditioned inverse problem in which different parameter combinations can produce similar signal decays(7,43). This limitation is reflected in the spatially heterogeneous quantification errors observed in **Figure S3.** For $R$ quantification, the largest RMSE occurred in the small-cell ($R < 4$ μm) and large-cell ($R > 10$ μm) regimes at low SNR, with errors remaining relatively elevated for large cells even at higher SNR. This likely reflects a mismatch between the diffusion times probed and the characteristic diffusion length scales of large cells. For $V_{in}$ quantification, estimation was most challenging at high $V_{in}$ (> 0.7) combined with small and large cell radii for both protocols, likely due to reduced contrast between intra- and extracellular diffusion components. Despite these intrinsic limitations, the optimized protocol consistently reduced RMSE compared with the baseline protocol.

4.5 In-Vivo Validation

In vivo results further demonstrated improved parameter map stability with the optimized protocol. The resulting maps exhibited smoother spatial patterns and fewer extreme outliers than those obtained with the baseline protocol, consistent with the expected spatial continuity of biological tissue properties. These improvements were achieved without increasing total acquisition time, supporting the feasibility of deploying the optimized protocol in clinical IMPULSED studies.

4.6 Generality and Limitations of the Framework

A key feature of our Bayesian experimental design framework is its foundation in information theory. Rather than optimizing task-specific metrics such as classification accuracy or RMSE for a particular fitting algorithm, the EIG objective directly quantifies the reduction in parameter uncertainty achievable from a given acquisition protocol. By maximizing the information content of acquired measurements, the optimized protocols support diverse analysis strategies, including classification, regression, and parameter estimation using different inference methods. Consequently, the optimization remains independent of downstream analysis methods. This flexibility is particularly valuable for IMPULSED diffusion MRI cytometry, where optimal quantification strategies remain an active area of investigation, and ongoing work includes neural network-based, traditional least-squares fitting, and Bayesian inference methods for parameter estimation. The framework also naturally incorporates nonlinear signal–parameter relationships and realistic noise models without requiring analytical tractability, since EIG is evaluated using Monte Carlo sampling over parameter priors and measurement noise. Further, although demonstrated here for the IMPULSED model, the same Bayesian optimization framework can be readily applied to other quantitative MRI techniques that rely on forward signal models linking tissue properties to measurements.

Several limitations should be acknowledged. First, the optimization assumes that the IMPULSED model accurately describes the underlying diffusion physics. Model inaccuracies—such as neglected membrane permeability, cellular shape anisotropy, or additional diffusion compartments—may bias the information content targeted by the optimization. Second, while the optimized protocols showed robustness to moderate prior misspecification, their performance may vary under substantially different biological conditions. Finally, although the in-vivo results are encouraging, ongoing longitudinal studies in this canine tumor model are needed. investigating changes in estimated parameters across multiple radiation treatment fractions. These studies will validate the method's sensitivity to treatment-induced microstructural changes and assess the optimized protocol's ability to detect clinically relevant biological alterations over time.

## 5. Conclusion

In this study, we developed and validated a Bayesian experimental design optimization framework for IMPULSED diffusion MRI cytometry. We demonstrated that information-theoretic optimization of pulse sequence parameters yields substantially more informative data acquisition, which improves both classification discriminability and parameter estimation accuracy compared with conventional acquisition strategies. Beyond establishing the practical benefits of optimized experimental design, our systematic analysis of optimization convergence under varying assumptions provides fundamental insights into IMPULSED model behavior under clinically relevant SNR constraints, providing actionable guidance for implementing diffusion MRI cytometry.

**Acknowledgments**

The authors thank Junzhong Xu, PhD, for valuable discussions and for sharing code used in this study.

**Tables**

| Protocol Type | Sequence | δ/Δ (ms) | $t_{\text{diff}}$ (ms) | Nonzero b-value (s/mm²) |
|---|---|---|---|---|
| Optimize SNR 20 | PGSE | 33/59 | 48 | 688(1), 800(1), 829(1), 931(1), 2000(4) |
|  | OGSEn1 | 41/51 | 10.25 | 1220(5) |
| Baseline | PGSE | 12/74 | 70 | 250(1), 500(1), 750(1), 1000(1), 1400(1), 1800(1) |
|  | OGSEn1 | 41/51 | 10.25 | 250(1), 500(1), 750(1), 750(1), 1000(1) |
|  | OGSEn2 | 41/51 | 5.125 | 100(1), 200(1), 300(1) |
| Baseline2 | PGSE | 33/59 | 48 | 250(1), 500(1), 750(1), 1000(1), 1400(1), 1800(1) |
|  | OGSEn1 | 41/51 | 10.25 | 250(1), 500(1), 750(1), 750(1), 1000(1) |
|  | OGSEn2 | 41/51 | 5.125 | 100(1), 200(1), 300(1) |
| Optimize SNR 40 | PGSE | 26/62 | 53.33 | 583(1), 710(1), 767(1), 871(1), 2000(3) |
|  | OGSEn1 | 41/51 | 10.25 | 1220(4) |
|  | OGSEn2 | 41/51 | 5.125 | 298(2) |

**Table 1: Comparison of optimized and baseline pulse sequence parameters.**
Optimized SNR 20 and Optimized SNR 40 protocol were obtained via Bayesian optimization maximizing Expected Information Gain under UR-UVin prior distribution with SNR=20 and SNR=40 noise assumptions respectively. The protocol applied by Xu et al. was used as Baseline protocol. While Baseline2 protocol was constructed based on Baseline by replacing the timing parameters with that of protocol Optimized SNR 20. All protocols contain 13 non-zero acquisitions. **Note:** The maximum OGSEn2 b-value (298 s/mm²) differs slightly from the baseline protocol (300 s/mm²) due to floating-point precision and minor variations in achievable gradient strength. δ: gradient pulse duration; Δ: time interval between two diffusion pulses; $t_{diff}$: effective diffusion time; PGSE: pulsed gradient spin echo; OGSE: oscillating gradient spin echo with oscillation number n. Numbers in parentheses indicate the number of acquisitions at the corresponding b-value.

| R CLASSIFICATION ACCURACY (%) | | | | |
|---|---|---|---|---|
| Protocol | SNR=5 | SNR=10 | SNR=20 | SNR=40 |
| Optimized SNR 20 | 39.70 ± 2.73 | 64.53 ± 2.59 | 91.90 ± 1.82 | 99.90 ± 0.15 |
| Optimized SNR 40 | 38.03 ± 2.87 | 60.27 ± 2.88 | 87.57 ± 1.20 | 99.10 ± 0.54 |
| Baseline | 31.30 ± 2.73 | 47.20 ± 1.69 | 74.03 ± 2.53 | 95.30 ± 1.06 |
| Baseline2 | 30.20 ± 3.02 | 47.80 ± 3.06 | 75.50 ± 1.80 | 96.43 ± 0.72 |

| VIN CLASSIFICATION ACCURACY (%) | | | | |
|---|---|---|---|---|
| Protocol | SNR=5 | SNR=10 | SNR=20 | SNR=40 |
| Optimized SNR 20 | 32.87 ± 2.77 | 50.17 ± 3.39 | 79.40 ± 2.94 | 98.47 ± 0.62 |
| Optimized SNR 40 | 28.77 ± 2.59 | 44.30 ± 3.41 | 72.47 ± 3.14 | 96.33 ± 1.04 |
| Baseline | 27.73 ± 2.61 | 39.30 ± 2.18 | 62.60 ± 2.40 | 90.30 ± 1.83 |
| Baseline2 | 28.27 ± 1.87 | 40.23 ± 2.80 | 64.60 ± 1.91 | 91.43 ± 1.57 |

**Table 2: Random Forest classification accuracy for cell population discrimination.**

Classification accuracy for distinguishing between two cell populations ($R$, $V_{in}$) using Random Forest classifier with 10-fold cross-validation. Training and testing performed on 100,000 simulated IMPULSED signals. Four protocols compared: Optimized SNR 20 (optimized for SNR=20 conditions), Optimized SNR 40 (optimized for SNR=10 conditions), Baseline and Baseline2 (timing setting set to that of Optimized SNR 20). Accuracy values represent mean performance across cross-validation folds. Higher values indicate better discrimination capability.

| Normalized RMSE of $R$ | | | | |
|---|---|---|---|---|
| Protocol | SNR=5 | SNR=10 | SNR=20 | SNR=40 |
| Optimized SNR 20 | 26.33 | 21.86 | 17.68 | 14.88 |
| Optimized SNR 40 | 25.77 | 21.32 | 17.50 | 15.12 |
| Baseline | 28.48 | 24.50 | 20.13 | 16.40 |
| Baseline2 | 28.30 | 24.48 | 20.12 | 16.41 |

| Normalized RMSE of $V_{in}$ | | | | |
|---|---|---|---|---|
| Protocol | SNR=5 | SNR=10 | SNR=20 | SNR=40 |
| Optimized SNR 20 | 21.59 | 17.73 | 14.82 | 13.14 |
| Optimized SNR 40 | 22.05 | 18.21 | 15.54 | 14.22 |
| Baseline | 24.27 | 19.39 | 16.01 | 14.25 |
| Baseline2 | 24.29 | 19.56 | 16.36 | 14.72 |

| Normalized RMSE of $D_{ex}$ | | | | |
|---|---|---|---|---|
| Protocol | SNR=5 | SNR=10 | SNR=20 | SNR=40 |
| Optimized SNR 20 | 24.86 | 19.38 | 13.21 | 9.30 |
| Optimized SNR 40 | 24.92 | 18.85 | 13.05 | 9.75 |
| Baseline | 26.22 | 21.58 | 15.56 | 10.75 |
| Baseline2 | 26.23 | 21.62 | 15.57 | 10.82 |

**Table 3: Quantification accuracy for biophysical parameters across protocols and noise levels**

Normalized root-mean-square error (RMSE) for 5-parameter least-squares fitting of IMPULSED model. RMSE values normalized by parameter upper bounds ($R$: 15 μm, $D_{in}$: 3 μm²/ms, $V_{in}$: 1.0, $D_{ex}$: 3 μm²/ms, $\beta_{ex}$: 10) and expressed as percentages for comparability across parameters. Lower values indicate more accurate parameter estimation. Quantification performed on 100,000 simulated signals with Rician noise at specified SNR levels using Trust Region Reflective algorithm.

| Prior SNR | Sequence | δ/Δ (ms) | $t_{\text{diff}}$ (ms) | b-value (s/mm²) |
|---|---|---|---|---|
| SNR 5 | PGSE | 34/58 | 46.67 | 1350(1), 1853(1), 1962(1), 2000(5) |
| | OGSEn1 | 41/51 | 10.25 | 1190(1), 1220(4) |
| | OGSEn2 | N/A | | |
| SNR 10 | PGSE | 32/60 | 49.33 | 798(1), 1004(1), 1316(1), 2000(4) |
| | OGSEn1 | 41/51 | 10.25 | 1105(1), 1109(1), 1194(1), 1220(3) |
| | OGSEn2 | N/A | | |
| SNR 20 | PGSE | 33/59 | 48.00 | 688(1), 800(1), 829(1), 931(1), 2000(4) |
| | OGSEn1 | 41/51 | 10.25 | 1220(5) |
| | OGSEn2 | N/A | | |
| SNR 40 | PGSE | 26/62 | 53.33 | 583(1), 710(1), 767(1), 871(1), 2000(3) |
| | OGSEn1 | 41/51 | 10.25 | 1220(4) |
| | OGSEn2 | 41/51 | 5.125 | 298(2) |

**Table 4: Pulse sequence parameters optimized for different noise conditions.**

Optimized protocol parameters obtained via Bayesian optimization under UR-UVin prior distribution at four SNR levels (5, 10, 20, 40). All optimizations used identical prior assumptions and scanner hardware constraints, varying only the noise level. Each optimization ran for 500 iterations to maximize Expected Information Gain. $\delta$: gradient pulse duration; $\Delta$, time interval between diffusion gradients; $t_{diff}$, effective diffusion time. Protocol adaptations reflect information-theoretic trade-offs between signal strength and diffusion weighting under different noise regimes. $\delta$: gradient pulse duration; $\Delta$: time interval between two diffusion pulses; $t_{diff}$: effective diffusion time; PGSE: pulsed gradient spin echo; OGSE: oscillating gradient spin echo with oscillation number n. Numbers in parentheses indicate the number of acquisitions at the corresponding b-value. **Note:** The maximum OGSEn2 b-value (298 s/mm²) differs slightly from the baseline protocol (300 s/mm²) due to floating-point precision and minor variations in achievable gradient strength.

| Prior Type | Sequence | δ/Δ (ms) | $t_{\text{diff}}$ (ms) | b-value (s/mm²) |
|---|---|---|---|---|
| UR-SVin | PGSE | 26/66 | 57.33 | 716(1), 745(1), 768(1), 916(1), 2000(4) |
| | OGSEn1 | 41/51 | 10.25 | 1220(5) |
| | OGSEn2 | N/A | | |
| UR-LVin | PGSE | 26/66 | 57.33 | 582(1), 749(1), 801(1), 848(1), 1905(1), 2000(3) |
| | OGSEn1 | 41/51 | 10.25 | 1124(1), 1220(4) |
| | OGSEn2 | N/A | | |
| SR-SVin | PGSE | 10/82 | 78.67 | 673(1), 683(1), 789(1), 1003(1), 2000(4) |
| | OGSEn1 | 41/51 | 10.25 | 1118(1), 1220(4) |
| | OGSEn2 | N/A | | |
| SR-UVin | PGSE | 10/82 | 78.67 | 694(1), 707(1), 819(1), 968(1), 977(1), 2000(3) |
| | OGSEn1 | 41/51 | 10.25 | 1154(1), 1220(3) |
| | OGSEn2 | 41/51 | 5.125 | 298(1) |
| SR-LVin | PGSE | 24/67 | 59.00 | 734(1), 832(1), 851(1), 1019(1), 2000(4) |
| | OGSEn1 | 41/51 | 10.25 | 1220(5) |
| | OGSEn2 | N/A | | |
| LR-SVin | PGSE | 25/60 | 51.77 | 705(1), 745(1), 765(1), 989(1), 2000(4) |
| | OGSEn1 | 40/52 | 10.25 | 934(1), 1004(1), 1093(1), 1098(1), 1131(1) |
| | OGSEn2 | N/A | | |
| LR-UVin | PGSE | 35/57 | 45.33 | 712(1), 753(1), 780(1), 841(1), 2000(4) |
| | OGSEn1 | 41/51 | 10.25 | 977(1), 1066(1), 1191(1), 1220(2) |
| | OGSEn2 | N/A | | |
| LR-LVin | PGSE | 29/63 | 53.34 | 754(1), 773(1), 790(1), 866(1), 1998(1), 2000(3) |
| | OGSEn1 | 40/52 | 10.25 | 1131(5) |
| | OGSEn2 | N/A | | |

**Table 5. Pulse sequence parameters optimized for different biological assumptions.** Optimized protocol parameters obtained via Bayesian optimization under SNR=20 noise assumption with nine different prior distributions over cell parameters. Prior distributions represent combinations of $R$ biases (Uniform, Small, Large) and $V_{in}$ biases (Uniform, Small, Large). UR-UVin, uniform $R$ and $V_{in}$; SR-UVin, small $R$-biased; LR-UVin, large $R$-biased; UR-SVin, small $V_{in}$-biased; etc. δ: gradient pulse duration; Δ: time interval between two diffusion pulses; $t_{diff}$: effective diffusion time; PGSE: pulsed gradient spin echo; OGSE: oscillating gradient spin echo with oscillation number n. Numbers in parentheses indicate the number of acquisitions at the corresponding b-value. **Note:** The maximum OGSEn2 b-value (298 s/mm²) differs slightly from the baseline protocol (300 s/mm²) due to floating-point precision and minor variations in achievable gradient strength.

**Figure Captions**

**Figure 1: Optimized protocols improve cell classification across noise conditions.**

Random Forest classification accuracy at SNR levels 5, 10, 20, and 40 for four protocols: *Optimized SNR 20* (blue), *Optimized SNR 40* (orange), *Baseline2* (green), and *Baseline* (grey). Optimized protocols achieve higher accuracy across all noise conditions, with greatest improvements at intermediate SNR.

**Figure 2: Optimized protocols reduce parameter estimation errors across noise conditions.**

Normalized RMSE (% of parameter maximum) for 3-parameter fitting at SNR 5, 10, 20, and 40. Optimized protocols (blue, orange) substantially reduce RMSE compared to Baseline (grey) and Baseline2 (green).

**Figure 3: Spatial distribution of quantification improvements from optimized protocols.**

Difference heatmaps showing RMSE improvement (Optimized SNR 20 - Baseline) across biologically relevant parameter space. Row 1 shows cell radius ($R$) quantification; Row 2 shows intracellular volume fraction ($V_{in}$) quantification. Columns represent SNR levels 5, 10, 20, and 40. Optimized protocol demonstrates improvements (predominantly blue) across nearly all biophysical regimes and noise conditions.

**Figure 4: In vivo comparison of baseline and optimized protocols for IMPULSED parameter estimation in canine tissue.**

Parameter maps derived from acquired in a canine model using two acquisition protocols. 1st row, 1~3 column: Results from the *Baseline* protocol. 2nd row, 1~3 column: Results from the SNR-*Optimized SNR 20* protocol. The estimated parameter maps are all overlaid on PGSE b=0 images within the tumor region of interest; Column 4 displays the full field-of-view PGSE diffusion-weighted images (top: b=0 s/mm², bottom: b=2000 s/mm²) with the red box indicating the zoomed region shown in columns 1-3; $R$: cell radius, $V_{in}$: intracellular volume fraction, $D_{ex}$: extracellular diffusivity; All parameter maps were estimated using nonlinear least-squares fitting with identical boundary constraints ($R: 0.2 - 15\ \mu m, V_{in}: 0 - 1, D_{ex}: 0 - 3.0\ \mu m^2/ms$). The optimized protocol yields visibly smoother parameter maps with reduced pixel-to-pixel variability and fewer extreme values, demonstrating improved estimation stability compared to the baseline protocol.

**Supplementary**

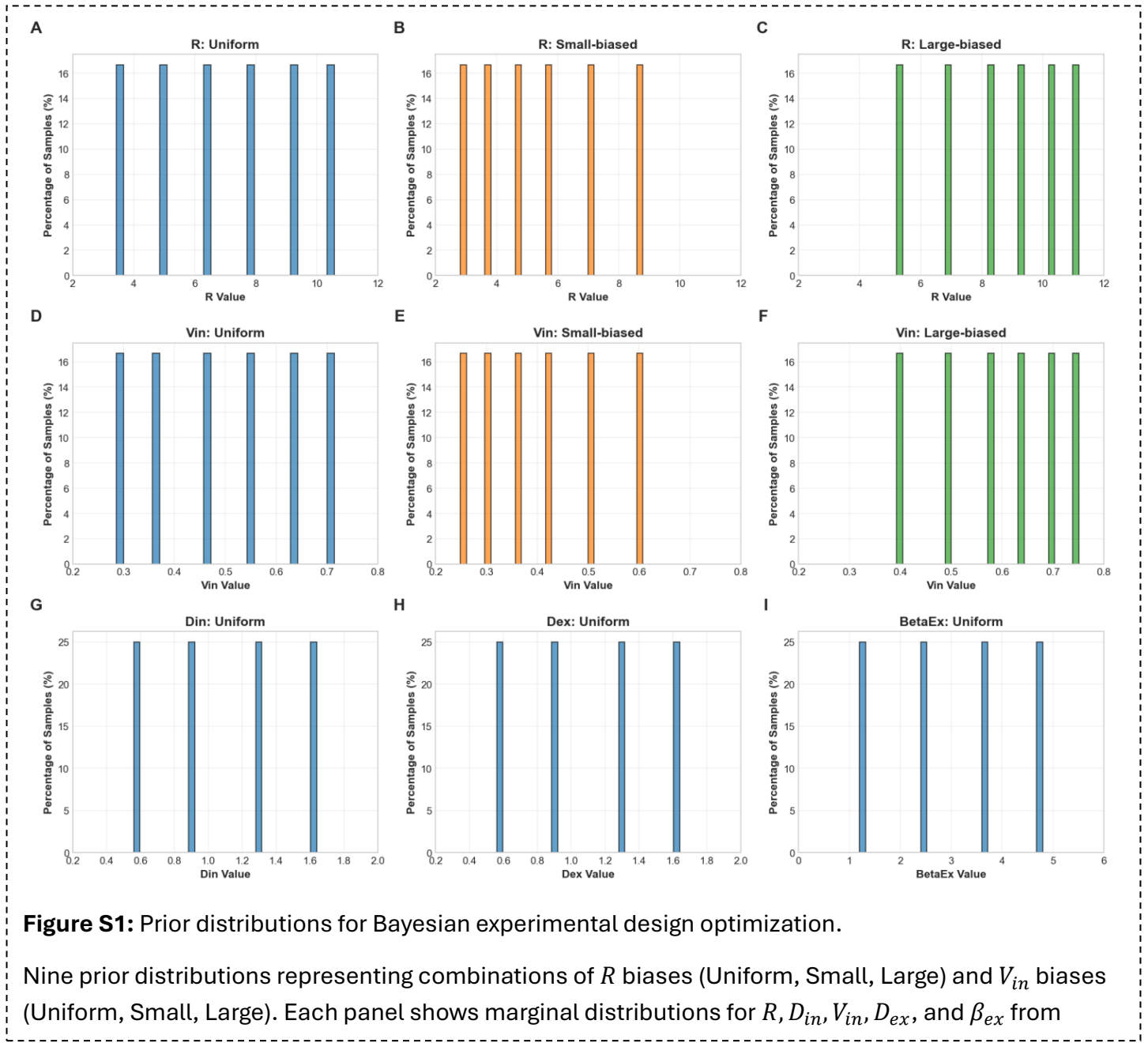

**Figure S1:** Prior distributions for Bayesian experimental design optimization.

Nine prior distributions representing combinations of $R$ biases (Uniform, Small, Large) and $V_{in}$ biases (Uniform, Small, Large). Each panel shows marginal distributions for $R, D_{in}, V_{in}, D_{ex}$, and $\beta_{ex}$ from

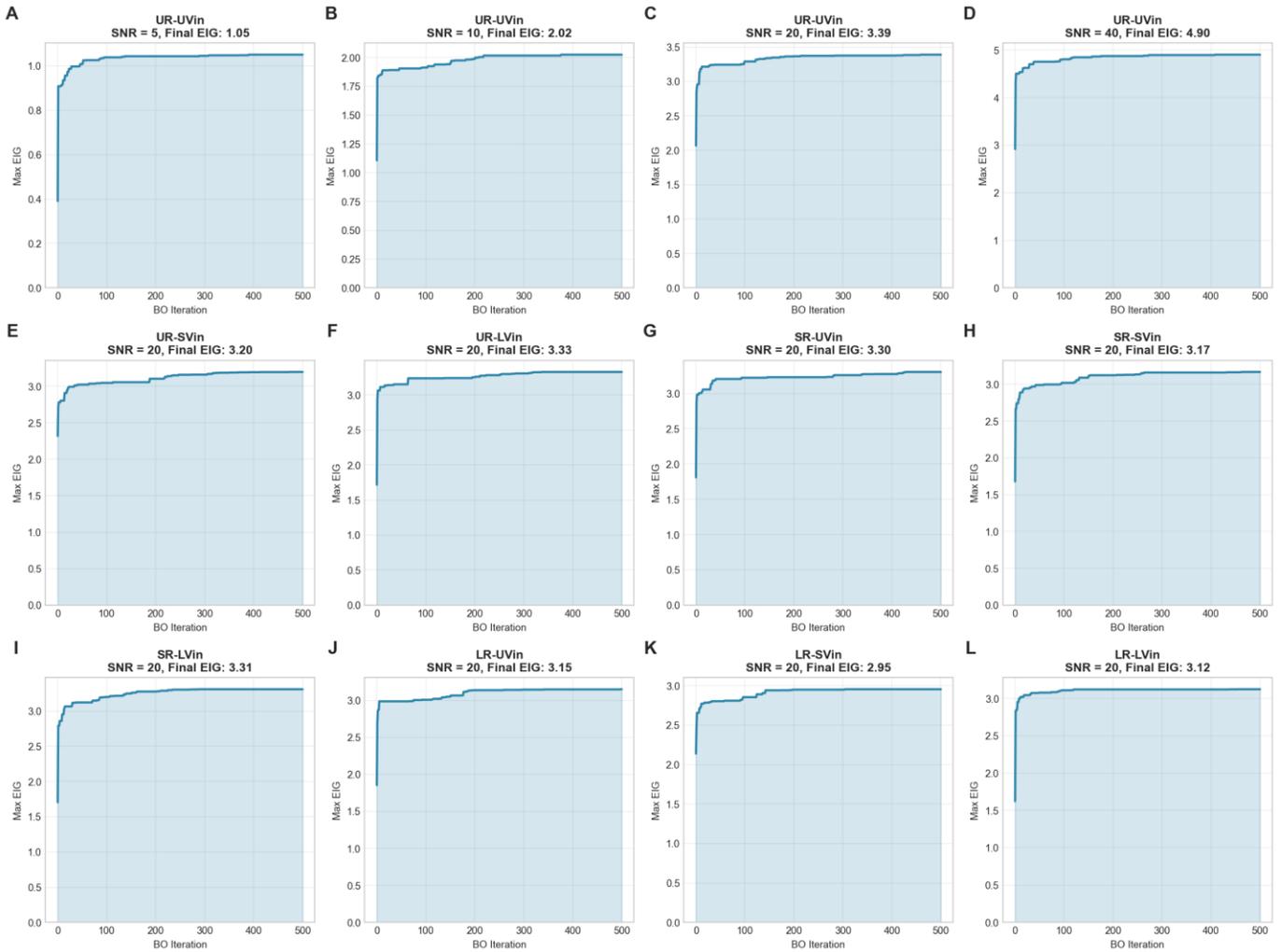

**Figure S2:** Convergence of Bayesian optimization across conditions.

Maximum EIG over 500 iterations for 12 optimization cases. Row 1 shows UR-UVin prior at SNR 5, 10, 20, 40, demonstrating strong SNR dependence (final EIG: 1.05 to 4.90). Rows 2-3 show eight additional priors at SNR=20. All optimizations converge within 100-200 iterations to similar EIG values (2.95-3.33), indicating noise level dominates over prior shape.

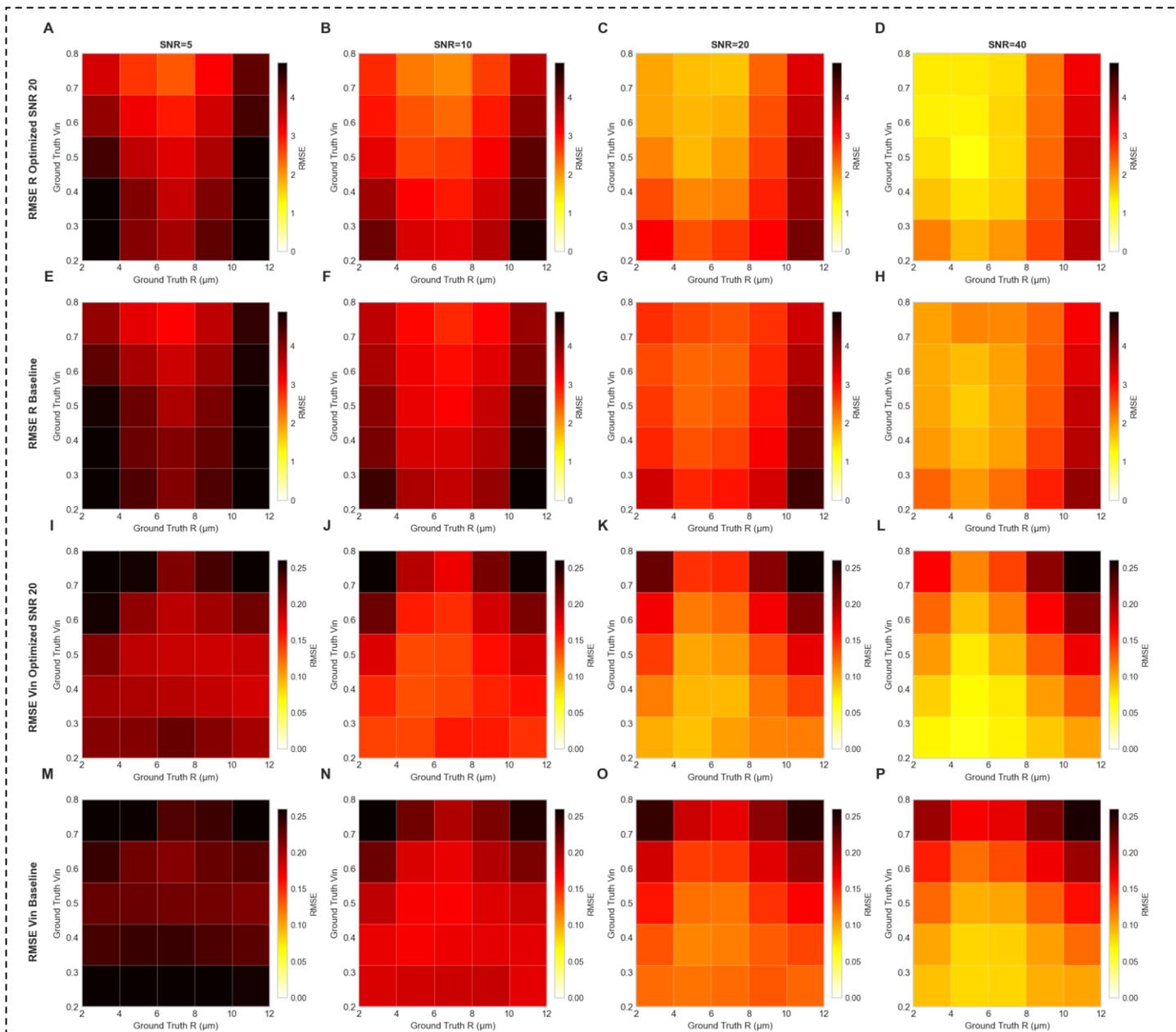

**Figure S3:** Direct comparison of absolute RMSE values between optimized and baseline protocols.

Absolute RMSE heatmaps for direct protocol comparison. Rows 1-2 show cell radius ($R$) quantification for Optimized SNR 20 and Baseline protocols, respectively. Rows 3-4 show intracellular volume fraction ($V_{in}$) quantification for Optimized SNR 20 and Baseline protocols, respectively. Columns represent SNR levels 5, 10, 20, and 40. Color intensity indicates normalized RMSE (% of parameter maximum). Comparison between row pairs reveals systematic RMSE reduction with optimized protocols across the parameter space.